\documentclass[aps,prl,amsmath,amssymb,twocolumn,superscriptaddress,preprintnumbers,floatfix,nofootinbib]{revtex4-2}

\usepackage{amsfonts}
\usepackage[utf8]{inputenc}
\usepackage{graphicx}
\usepackage{comment}
\usepackage{cancel}
\usepackage{hyperref}
\usepackage{physics}
\usepackage{slashed}
\usepackage{enumitem}
\usepackage{appendix}
\usepackage[usenames,dvipsnames]{xcolor}
\usepackage{url}
\usepackage[normalem]{ulem}
\usepackage{xcolor}

\newcommand{\mc}[1]{\mathcal{#1}}

\newcommand{\ra}{\rightarrow}

\newcommand{\bb}{\bar{b}}
\newcommand{\svbsf}{\left \langle \sigma v \right \rangle_\mathcal{B}}
\newcommand{\GB}{ \left \langle \Gamma_\mc{B} \right \rangle}

\newcommand{\oo}[1]{\textcolor{olive}{#1}}

\hypersetup{pdfstartview=FitV,colorlinks=true,linkcolor=blue,citecolor=red,filecolor=black,urlcolor=blue,breaklinks=true}

\usepackage[parfill]{parskip}
\begin{document}\sloppy 

\preprint{MITP-24-067}
\preprint{KEK-TH-2642}

\vspace*{1mm}

\title{Impact of Bound State Formation on Baryogenesis}

\author{Mathias Becker}
\email{bmathias@uni-mainz.de}
 \affiliation{PRISMA$^+$ Cluster of Excellence \& Mainz Institute for Theoretical Physics,\\ FB 08 - Physics, Mathematics and Computer
Science,\\ 
 Johannes Gutenberg-Universit\"{a}t Mainz, 55099 Mainz, Germany} 
\author{K\aa re Fridell}
\email{kare.fridell@kek.jp}
 \affiliation{Theory Center, Institute of Particle and Nuclear Studies, High Energy Accelerator Research Organization (KEK), Tsukuba 305-0801, Japan}
 \affiliation{Department of Physics, Florida State University, Tallahassee, FL 32306, USA}
\author{Julia Harz}
\email{julia.harz@uni-mainz.de}
\affiliation{PRISMA$^+$ Cluster of Excellence \& Mainz Institute for Theoretical Physics,\\ FB 08 - Physics, Mathematics and Computer
Science,\\ 
 Johannes Gutenberg-Universit\"{a}t Mainz, 55099 Mainz, Germany}
\author{Chandan Hati}
\email{chandan@ific.uv.es}
\affiliation{Instituto de F\'{i}sica Corpuscular (IFIC), Universitat de Val\`encia-CSIC,\\C/ Catedratico Jose Beltran, 2, E-46980 Valencia, Spain}
\vspace{0.5cm}

\begin{abstract} 
The mechanism behind the generation of the baryon asymmetry of the Universe (BAU) is one of the biggest open questions of (astro-)particle physics. Popular mechanisms to generate the observed baryon asymmetry include CP-violating out-of-equilibrium decays and scatterings of heavy particles. If these heavy non-relativistic particles feature long-range interactions, the formation of bound states can impact the generation of the baryon asymmetry. We outline the general conditions for when bound states are important for decay and scattering dominated baryogenesis and present the necessary Boltzmann equations for the first time.
We demonstrate that bound states can impact baryogenesis in three different ways: They (i) strongly impact abundances of particles sourcing the BAU, (ii) act as a source term of the asymmetry, and (iii) mediate additional washout channels.

\end{abstract}

\maketitle

\setcounter{equation}{0}

\textbf{Introduction---}
The observed baryon asymmetry of the Universe (BAU) cannot be explained within the Standard Model (SM) of particle physics and requires the existence of new Physics.
The three Sakharov conditions required to generate a baryon asymmetry are the violation of charge ($C$) and charge-parity ($CP$) symmetries, the existence of baryon number violating (BNV) processes, as well as interactions out of equilibrium. 
A common way to realize these conditions is to introduce additional baryon ($B$) or lepton ($L$) number violating interactions, where an asymmetry can be generated by either decay~\cite{Fukugita:1986hr} or scattering~\cite{Yoshimura:1978ex,Kolb:1979qa,Kolb:1990vq} of some new heavy out-of-equilibrium state. 
If the decaying or scattering field(s) are charged under the SM (or a new dark) gauge group with coupling strength $\alpha$, exchanges of gauge bosons with relatively light masses (compared to the heavy fields) can alter the evolution of particle densities when $\alpha \sim v$, where $v$ is the relative velocity between a pair of heavy particles. This affects the size of the BAU nontrivially, e.g.\ by enforcing thermal equilibrium for a longer period. The same arguments apply to Yukawa or scalar interactions.
Interestingly, a vast number of popular scenarios for baryogenesis feature these possibilities, for instance, decay based mechanisms in the case of type-II seesaw leptogenesis~\cite{Ma:1998dx,Lazarides:1998iq,Hambye:2000ui,Hambye:2003ka,Hambye:2005tk, GonzalezFelipe:2013jkc, AristizabalSierra:2014nzr}, type-III seesaw leptogenesis~\cite{Hambye:2003rt,Fischler:2008xm,AristizabalSierra:2010mv,Hambye:2012fh}, or scattering-based mechanisms~\cite{Bento:2001rc,Baldes:2014gca,Farrar:2005zd,Gu:2009yx,Blennow:2010qp,Cui:2011ab,Kumar:2013uca,Dasgupta:2016odo,Borah:2018uci,Dasgupta:2019lha}.
If the above conditions are fulfilled, non-perturbative effects---the Sommerfeld effect and the formation of bound states---play a crucial role in the accurate prediction of the viable parameter space. 
This is a well-known effect in the context of dark matter (DM) relic abundance calculations~\cite{Ellis:2015vaa,Harz:2018csl,Garny:2021qsr,Biondini:2018ovz,Becker:2022iso,Binder:2023ckj}. 
The Sommerfeld effect for leptogenesis has been discussed in the context of a type-II/III seesaw~\cite{Strumia:2008cf}. 
In this Letter, we study for the first time how the existence of bound states can affect the evolution of particle densities and asymmetry generation in the cases of decay and scattering dominated baryogenesis mechanisms. 
We find that bound states can: (i) strongly impact abundances of particles sourcing the BAU, (ii) act as a source term of the asymmetry, (iii) mediate additional washout channels. Our findings show that bound states can reduce the baryon asymmetry in decay dominated baryogenesis by more than an order of magnitude, while in scattering dominated scenarios, bound states can enhance or decrease the asymmetry by $\mc{O} (1)$ factors.
%%%%%%%%%%%%%%%%%%%%%%%%%%%%%%%%%%%%%%%%%%%%%%%%%%%%%%%%%%
%%%%%%%%%%%%%%%%%%%%%%%%%%%%%%%%%%%%%%%%%%%%%%%%%%%%%%%%%%
%%%%%%%%%%%%%%%%%%%%%%%%%%%%%%%%%%%%%%%%%%%%%%%%%%%%%%%%%%

\textbf{Bound state formation and the steady state approximation---}
If two particles $X$ charged under a gauge group start to become non-relativistic, they can efficiently form bound states $\mc{B} \left( XX \right)$ by the emission of a gauge boson ($V$), 
   $ XX \ra \mathcal{B} \left( XX \right) V$,  
a process we will refer to as bound state formation (BSF). 
We note that our discussion can be straightforwardly generalized to any interaction of mediators being light compared to $X$~\cite{Oncala:2018bvl,Petraki:2016cnz}.
For simplicity, we will focus on the example of gauge interactions and consider the particle, responsible for the asymmetry generation, to be charged under a non-abelian gauge group. 
Concretely, we consider an example where $X$ transforms in the adjoint representation of the non-abelian gauge group $SU(N_D)$~\footnote{The analysis can be straightforwardly generalised for heavy particles transforming under any other representation of $SU(N_D)$ leading to BSF. When numerical results are presented in the following, we choose $N_D=2$.}. The BSF cross section for the ground state is then given by~\cite{Petraki:2015hla,Harz:2018csl}
\begin{align}
\hspace{-0.18cm} \sigma_{\mc{B}}v = \frac{2^9 \pi}{3} \frac{\alpha^\text{BSF} \alpha_g^B}{m_X^2} \frac{ N_D^3 + 4 N_D}{8\left( N_D^2 + 1 \right)^2} S_\mc{B} \left( \frac{\alpha_g^S}{v}, \frac{\alpha_g^B}{v} \right) \, , \label{eq:BSF_CS} 
\end{align}
which describes BSF from an initial state in the adjoint representation into a singlet bound state~\footnote{We neglect the contributions from BSF to excited states or more weakly bound states in higher representations of $SU(N_D)$.}. 
Here, $\alpha^\text{BSF}$ as well as $\alpha_g^B$, $\alpha_g^S$ are related to the gauge coupling $\alpha$ evaluated at different energy scales~\footnote{The superscript indicates at which energy scale the coupling constant is evaluated: $B$ refers to the typical energy exchange between the constituents of the bound state (the Bohr momentum), $S$ refers to the typical energy exchange between the particles in the scattering state (the relative momentum) and BSF is identified with the average energy of the emitted gauge boson. The subscript $g$ indicates that $\alpha$ should be multiplied by the appropriate group factor for the effective potential of the scattering state or bound state considered.}, and $S_\mc{B} \left( \alpha_g^S/v, \alpha_g^B/v \right)$ is a Sommerfeld factor, see e.g.\ Ref.~\cite{Harz:2018csl} for details. 
From studies of BSF in the context of DM~\cite{Ellis:2015vaa,Harz:2018csl,Garny:2021qsr,Biondini:2018ovz}, it is known that the bound state enters the Boltzmann equations as a new degree of freedom. Therefore the evolution of its number density $n_\mc{B}$ must be taken into account. In practice, however, the system can be effectively simplified by employing the steady state approximation \cite{Ellis:2015vaa,Binder:2021vfo}, which implies that the interaction rates of the bound states (either BSF and ionization or their decay processes) are large compared to the Hubble rate such that one can approximate
 $   \frac{\dd Y_\mathcal{B}}{\dd z} \approx 0$, 
where $z=m_X/T$, is a time variable and $Y_i\equiv n_i/s$, with $s$ denoting the entropy density.
This allows one to express the ratio $Y_\mc{B} / Y_\mc{B}^\text{eq}$ in terms of the bound state interaction rates and effectively encode the bound state effects in the remaining evolution equations. 
In what follows, we will illustrate the impact of BSF on the time evolution of asymmetries in the early universe. We will employ two simplified model examples representing decay and scattering dominated scenarios, before commenting on various realizations. 

%%%%%%%%%%%%%%%%%%%%%%%%%%%%%%%%%%%%%%%%%%%%%%%%%%%%%%%%%%

\textbf{Decay dominated Baryogenesis---}
To discuss the impact of BSF in the decay dominated asymmetry generation scenario, let us consider a simple model with a scalar $X$ (an adjoint of a non-abelian gauge group $SU(N_D)$ with a coupling strength $g$, $\alpha = g^2/(4 \pi)$ to an associated massless vector boson $V$) that can undergo $CP$ and $(B-L)$ violating decays into scalar final states $bb$ and $\bar{b}\bar{b}$ (with $b$ carrying a non-zero baryon number), such that the amplitude squared can be expressed in terms of a dimensionless BNV and CPV coupling $\lambda$ as
\begin{align}
    |\mc{M} \left( X \ra b b  \right)|^2 &= \lambda^2 m_X^2 \left( 1 + \epsilon \right) \, , \nonumber\\
    |\mc{M} \left( X \ra \bb \bb  \right)|^2 &= \lambda^2 m_X^2 \left( 1 - \epsilon \right) \, ,
\end{align}
where $\epsilon$ parameterizes the asymmetry of the decay.  
The standard Boltzmann equations for the asymmetry evolution in such a scenario take into account the BNV decays $X \rightarrow bb / \bb \bb$, scalar annihilations into gauge bosons $X X \rightarrow V V$ and the $X$ mediated washout process $bb \leftrightarrow \bb \bb$. 
In the presence of BSF of two $X$'s mediated by the gauge interaction, one must take into account BSF following Eq.~\eqref{eq:BSF_CS}, the inverse process of bound state ionization, and any additional bound state decay modes. 
The latter proceeds via the annihilation of the constituents of the bound state ($XX$) into mediators ($VV$) or $b\bar{b}$ with the dominant mode being the decay of $\mathcal{B}$ into two gauge bosons $(XX\rightarrow \mathcal{B} V) \rightarrow VVV$~\footnote{We find that bound state effects are mainly relevant for $\lambda \ll \alpha$, implying that the $\mc{B} \rightarrow b \bb$ channel is subdominant.}. 
We further note that since all of these processes are mediated by gauge interactions, they do not lead to any new sources of $CP$ or $(B-L)$ violation. The steady-state approximation then implies
\begin{align}
     \svbsf \left( Y_X^2 - \frac{Y_{\mc{B}}}{Y_{\mc{B}}^\text{eq}} {Y_X^\text{eq}}^2 \right) + \left \langle \Gamma_\mc{B} \right \rangle \left( Y_\mc{B} -Y_\mc{B}^\text{eq} \right)\approx 0 \, ,
\end{align}
which can be readily employed to simplify the evolution of the abundances $Y_X$ and $Y_{\Delta b} = Y_b - Y_{\bb}$ \footnote{
Before the electroweak phase transition, $\dot{Y}^{\text{sph}}_{\Delta(B-L)} =0$ is conserved due to the (B+L)-violating electroweak sphalerons.
In the presence of any additional BNV interactions $\dot{Y}^{\text{new}}_{\Delta B}\neq 0$, one can write $ \dot{Y}_{\Delta(B-L)} =  \dot{Y}^{\text{sph}}_{\Delta(B-L)} +  \dot{Y}^{\text{new}}_{\Delta(B-L)} =  \dot{Y}^{\text{new}}_{\Delta B}\, $.
Using the chemical potential relations between the SM degrees of freedom~\cite{Harvey:1990qw}, the final BAU is given by $\frac{79}{28}\dot{Y}_{\Delta B} =  \dot{Y}^{\text{new}}_{\Delta B}$. We note that the numerical pre-factor drops out in the ratio of the final BAU with and without BSF.  ${Y}_{\Delta B}$ is related to $Y_{\Delta b}$ by the baryon number of $b$.

},
\begin{align}
      &c \frac{\dd Y_X}{\dd z} = -\frac{ \left \langle \Gamma_X \right \rangle}{z^2s} \left({Y_X} - Y_X^\text{eq} \right) - \frac{2 \left \langle {\sigma v}  \right \rangle_{\scriptscriptstyle{XX \rightarrow VV}}^\text{eff}}{z^2} \left( {Y_X^2} - {Y_X^\text{eq}}^2 \right) \, , %\label{eq:DecayBZ1}
      \nonumber\\
&c \frac{\dd Y_{\Delta b}}{\dd z} = \epsilon \frac{\left \langle \Gamma_X \right \rangle}{z^2 s} \left( Y_X - Y_X^\text{eq} \right) - 2\frac{Y_{\Delta b} Y_b^\text{eq}}{z^2} \left \langle {\sigma v} \right \rangle_{\scriptscriptstyle bb \leftrightarrow \bb \bb}  \, . \label{eq:DecayBZ2}  
\end{align}
Here $c = H/sz$ with the Hubble rate defined as $H\equiv\sqrt{\frac{8\pi^3 g_{*}}{90}} \frac{T^2}{m_{\text{Pl}}}$, where $g_{*}$ is the effective number of degrees of freedom, $m_{\text{Pl}}$ is the Planck mass and $Y_i (Y_i^\text{eq})$ is the yield (equilibrium yield) of particle species $i$. In deriving~Eq.\eqref{eq:DecayBZ2}, we have taken into account the real-intermediate state (RIS) subtraction~\cite{Giudice:2003jh}. The effective $X$ annihilation cross section including BSF effects is given by

\begin{align}
    \left \langle \sigma v \right \rangle_{\scriptscriptstyle XX \rightarrow VV}^\text{eff} &= \left \langle \sigma v \right \rangle_{\scriptscriptstyle XX \rightarrow VV} + \left \langle \sigma v \right \rangle_\mathcal{B}^\text{eff}   \, , \label{eq:effective_annihilation}
\end{align}
where $\Gamma_\mc{B}$ and $\Gamma_\text{ion}$ are the rates of bound state decay and ionization, respectively and $\left \langle \sigma v \right \rangle_\mathcal{B}^\text{eff} = \frac{\GB}{\GB + \left \langle \Gamma_\text{ion} \right \rangle} \svbsf$.
In this scenario, as the bound states are formed and decay in a baryon number conserving way, they only impact the $X$ evolution equation.
The formation of bound states and the subsequent decay ($XX \rightarrow \mathcal{B} V \rightarrow VVV$) efficiently depletes the $X$ number density, cf.\ Eq.~\eqref{eq:effective_annihilation}, which sources the baryon asymmetry.
Consequently, BSF leads to a reduction of the created asymmetry. We label the point in time from which BSF constitutes the dominant contribution to Eq.~\eqref{eq:effective_annihilation} as $z_\text{BSF}$, which we estimate to be $z_\text{BSF} \sim 0.2 \alpha^{-2}$~\footnote{The parametric form of this inequality can be obtained by assuming that bound states are efficiently formed when the ionisation of bound states becomes inefficient. This occurs when the temperature drops below the binding energy $E_\mc{B} \sim m_X \alpha^2$, such that the efficiency factor in \eqref{eq:effective_annihilation} is maximized. The numerical prefactor reflects the result of our numerical analysis.}.  
The asymmetry generation by $X$ decays approximately takes place at $z_\text{dec} \sim (10^{-7}/\lambda) \sqrt{m_X/\mathrm{TeV}}$.
We expect large effects from BSF if the asymmetry is generated late, $z_\text{dec}>z_\text{BSF}$, defining a \emph{late-decay scenario}. 
For the opposite hierarchy, $z_\text{BSF}>z_\text{dec}$ an \emph{early-decay scenario}, we expect BSF to be less relevant. 

\begin{figure}[t]
   \centering
    \includegraphics[width=\columnwidth]{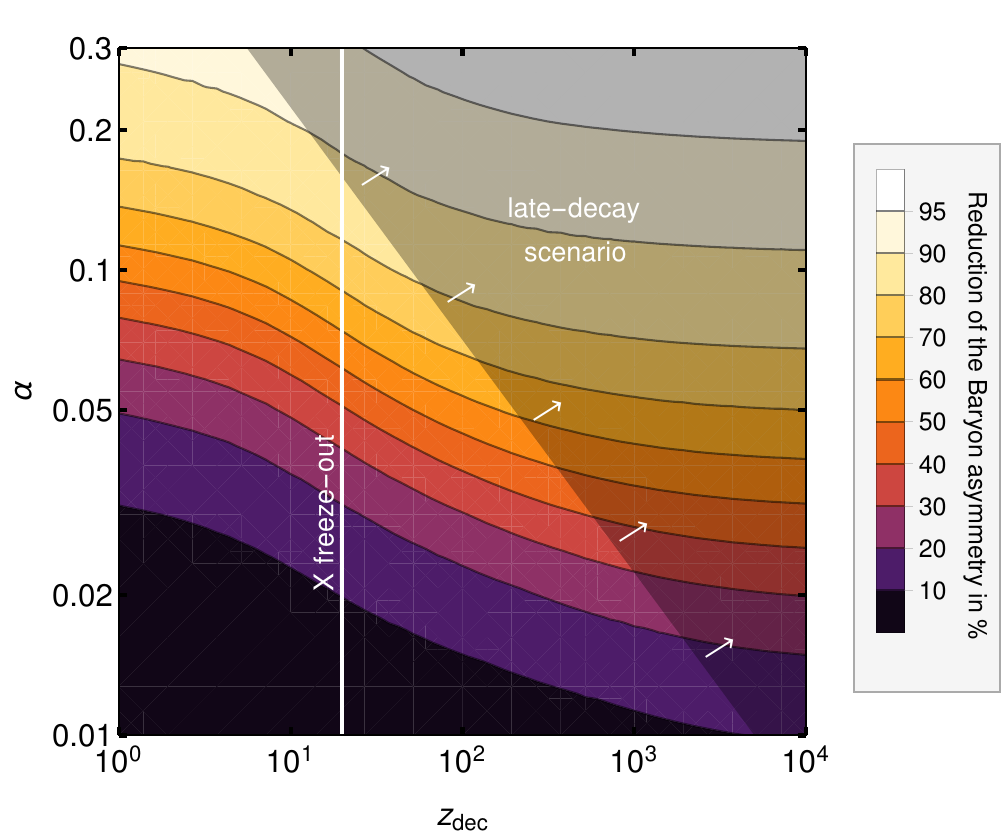}
    \caption{{\bf Decay dominated baryogenesis scenario.}
    The color code indicates the reduction of the baryon asymmetry in percent when including BSF. The results are shown in the $\alpha$ vs.~$z_\text{dec}$ plane for the benchmark choice $m_X=10^6$ GeV. In the shaded region, BSF dominates over perturbative annihilations at the time of asymmetry production, $z_\text{BSF}<z_\text{dec}$. See text for more details.
    }
    \label{fig:Decay_comp}
\end{figure}

To quantify the effects of BSF numerically, we integrate the Boltzmann equations in Eqs.~\eqref{eq:DecayBZ2} with and without considering bound state effects.
Taking the benchmark choice\footnote{Note that the results presented in the following are fairly independent of the actual choice of $m_X$ and mainly depend on $z_\text{dec}$. Thus, for the parameter points that do not reproduce the observed BAU in Fig.~\ref{fig:Decay_comp}, typically another $m_X$ can be found that leads to the correct BAU.
} $\epsilon=0.1$, $m_b=0.1 m_X$, $m_X = 10^6 \, \mathrm{GeV}$ and varying $\lambda$ and $\alpha$, we show in Fig.~\ref{fig:Decay_comp} the reduction of the baryon asymmetry when including BSF in terms of the gauge coupling $\alpha$ and the estimated time of decay $z_\text{dec}$.
%%%%%%%%%%%%%%%%%%%%%%%%%%
%%%%%%%%%%%%%%%%%%%%%%%%%%
In an \emph{early-decay} scenario, we find $\mc{O} (1-10) \%$ corrections.
Note that those deviations are induced mostly by the Sommerfeld corrections to the perturbative annihilation cross section, not by BSF.
On the contrary, for a \emph{late-decay} scenario, the presence of bound states significantly reduces the $X$ abundance before their decay and therefore also reduces the asymmetry.
The amount of reduction increases with the strength of the gauge interaction $\alpha$ and the time scale of the decay $z_\text{dec}$, and can be larger than one order of magnitude ($>90 \%$).
This late-decay scenario resembles the case of Super-WIMP DM production~\cite{Covi:1999ty,Feng:2003uy,Bollig:2021psb}, where decays of a heavy parent particle occur after it freezes out and bound states~\footnote{Excited bound states become more and more relevant for later decays of the dark matter parent particle. This will also be the case in the baryogenesis scenario discussed here. While we omit excited states in this analysis for simplicity, we expect them to lead to an increased reduction of the final asymmetry for large $z_\text{dec}$. For instance, in Super-WIMP DM scenarios, the final DM yield can be reduced by $\mc{O}(1\%-10\%)$ for $z_\text{dec} \sim \mc{O} (10^3 - 10^4)$ and a hundred excited states~\cite{Binder:2023ckj}.} can cause a significant reduction of the relic abundance \cite{Bollig:2021psb,Biondini:2020ric}.

%%%%%%%%%%%%%%%%%%%%%%%%%%%

\begin{figure}
    \centering
    \includegraphics[width=\columnwidth]{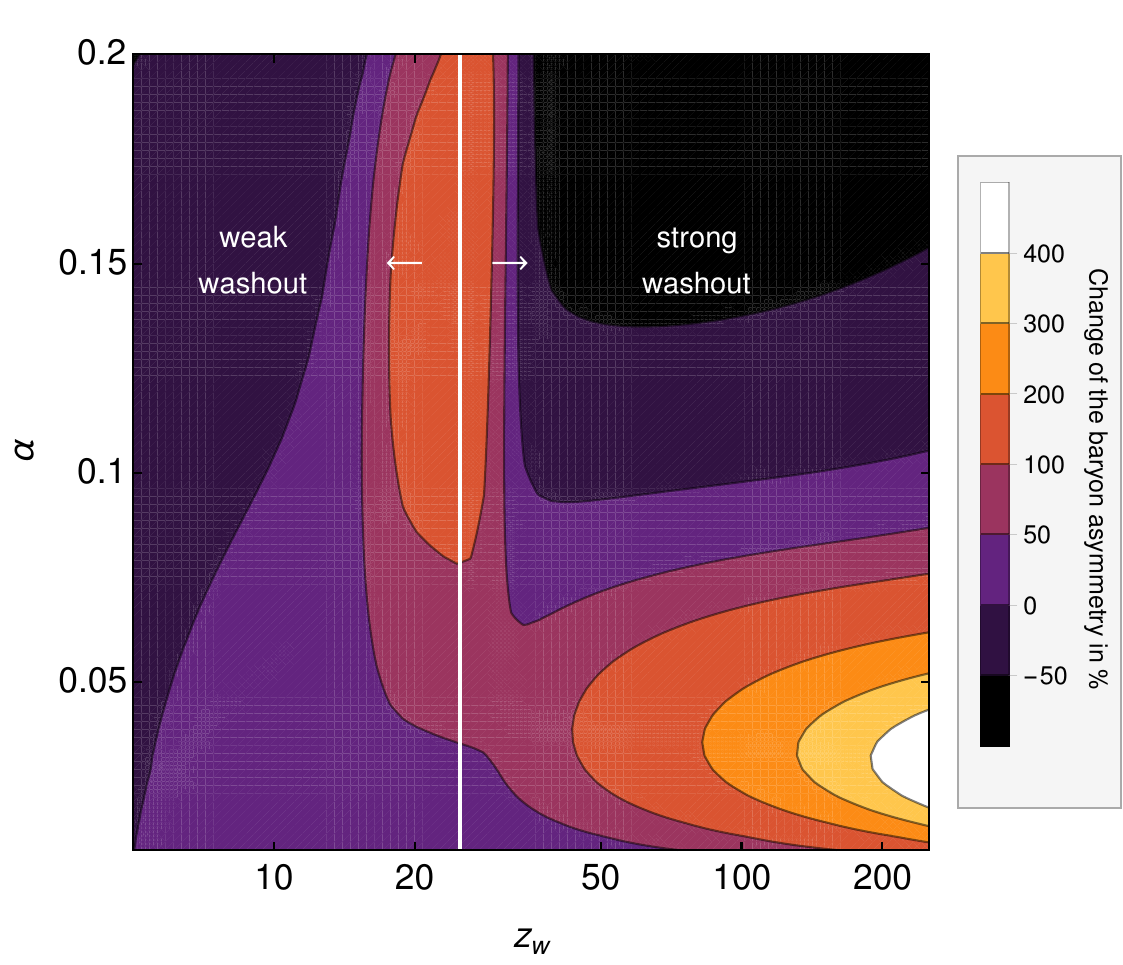}
    \caption{{\bf Scattering dominated baryogenesis scenario.} The change of the generated asymmetry in percent  when considering BSF in the scattering dominated scenario is illustrated in the $z_\text{w}$-$\alpha$ plane for $m_\phi=10^4 \, \mathrm{GeV}$, $\lambda_1=0.1$. The white line indicates the transition from the weak to the strong washout regime.}
    \label{fig:Scat_comp1}
\end{figure}

\textbf{Scattering dominated Baryogenesis}---
To discuss the impact of BSF on the scattering dominated asymmetry generation scenario, let us consider a model where a baryon asymmetry is generated via the $CP$- and $(B-L)$-violating scatterings of a pair of scalars $\phi$, again transforming in the adjoint representation of $SU(N_D)$
\begin{align}
    |\mathcal{M} \left( \phi \phi \rightarrow b b \right) |^2 &= |\lambda_1|^2 \left( 1+\epsilon \right) \, , \nonumber\\
    |\mathcal{M} \left( \phi \phi \rightarrow \bar{b} \bar{b} \right) |^2 &= |\lambda_1|^2 \left( 1-\epsilon \right) \, ,
\end{align}
where $\lambda_1$ is the dimensionless quartic coupling, $\epsilon$ parametrizes the asymmetry of the annihilation.
CPT invariance and unitarity further necessitate the existence of another CP-violating channel for the $bb$/$\bb \bb$ system, which we choose to be mediated via another dimensionless quartic coupling $\lambda_2$,
\begin{align}
     |\mathcal{M} \left( b b \rightarrow \bar{b} \bar{b} \right) |^2 &= |\lambda_2|^2 \left( 1+ \frac{|\lambda_1|^2}{|\lambda_2|^2}\epsilon \right)  \, , \label{eq:l2scattering1} \\
    |\mathcal{M} \left( \bar{b} \bar{b} \rightarrow b b \right) |^2 &= |\lambda_2|^2 \left( 1- \frac{|\lambda_1|^2}{|\lambda_2|^2}\epsilon \right) \, . \label{eq:l2scattering2}
\end{align}
The bound states formed by a pair of $\phi$'s can decay via the annihilation channels of the constituents $\phi$, which at leading order implies the existence of three possible decay modes ($\phi\phi\rightarrow \mc{B})\rightarrow$ $VV$, $bb$ and $\bb \bb$. Since the decay width of a bound state in the ground state is directly proportional to the $s$-wave annihilation cross section of the constituents, the decays of bound states $\mc{B} (\phi \phi)$ obey $\frac{\text{Br}_b-\text{Br}_{\bb}}{\text{Br}_b+\text{Br}_{\bb}} = \epsilon$, with $\text{Br}_i = \frac{\Gamma (\mc{B} \rightarrow ii)}{\Gamma_\mc{B}}$, and $\Gamma_\mc{B}$ being the total decay width of the bound state. The steady-state approximation in this scenario leads to
\begin{align}
      \svbsf \hspace{-0.1 cm} \left( Y_{\phi}^2 - \frac{Y_\mc{B} {Y_\phi^\text{eq}} ^2 }{Y_\mc{B}^\text{eq}}  \right) \hspace{-0.1 cm} +\hspace{-0.4 cm} \sum_{i=\left \lbrace V, b, \bb \right \rbrace} \hspace{-0.4 cm} \left \langle \Gamma_\mc{B} \right \rangle \text{Br}_i \hspace{-0.1 cm} \left( Y_\mc{B} -  \frac{Y_\mc{B}^\text{eq}{Y_i}^2}{{Y_i^\text{eq}}^2} \right) \approx 0 \, ,\nonumber
\end{align}
which simplifies the evolution of the abundances,
\begin{align}
c \frac{\dd Y_\phi}{\dd z} &= - \frac{2 }{z^2} \left[ \left \langle \sigma v \right \rangle_{\scriptscriptstyle \phi \phi \rightarrow bb}^\text{tot} +  \left \langle \sigma v \right \rangle_{\scriptscriptstyle \phi \phi \rightarrow VV}^\text{eff} \right] \left( Y_\phi^2 - \left( Y_\phi^\text{eq} \right)^2 \right) \, , \nonumber\\
 c \frac{\dd Y_{\Delta b}}{\dd z} &=  \frac{2}{z^2} \epsilon \left[ \left \langle \sigma v \right \rangle^\text{tot}_{\scriptscriptstyle \phi \phi \rightarrow bb} + \left \langle \sigma v \right \rangle_\mc{B}^\text{asy}  \right]\left( Y_\phi^2 - \left( Y_\phi^\text{eq} \right)^2 \right) \nonumber \\ &- \frac{Y_{\Delta b}}{Y_b^\text{eq}z^2} \left[ \left( Y_\phi^\text{eq}\right)^2\left \langle \sigma v \right \rangle^\text{tot}_{\scriptscriptstyle \phi \phi \rightarrow bb} + 2\left( Y_b^\text{eq}\right)^2\left \langle \sigma v \right \rangle_{\scriptscriptstyle bb \leftrightarrow \bb \bb}^\text{tot} \right. \nonumber \\ &\left. + \left( Y_b^\text{eq}\right)^2\left \langle \sigma v \right \rangle_{\scriptscriptstyle bb \leftrightarrow VV}^\text{tot} \right] \, , \label{eq:BoltzmannDeltaScat}  
\end{align}
where $\sigma^\text{tot}_{\phi \phi \to bb} = \sigma_{\phi \phi \rightarrow bb} + \sigma_{\phi \phi \rightarrow \bb \bb}$, $\left \langle \sigma v \right \rangle_{\phi \phi \rightarrow VV}^\text{eff}$ is given by Eq.~\eqref{eq:effective_annihilation} but with $X$ replaced by $\phi$, and $\sigma_{bb \leftrightarrow VV}^\text{tot} =  \sigma_{bb \leftrightarrow VV} +  \sigma_{\bb \bb \leftrightarrow VV}$. 
Similar as in the decay scenario, RIS is taken into account for the bound state mediated scattering.
We can readily infer three distinct and important effects due to bound states in the case of scattering dominated baryogenesis: Firstly, as for the decay dominated baryogenesis scenario, the bound states increase the annihilation cross section of the particles sourcing the asymmetry, resulting in a depletion of the generated asymmetry. Secondly, due to their $CP$- and $(B-L)$-violating decay modes in this scenario, the bound states also induce an asymmetry generating term 
\begin{align}
 \left \langle \sigma v \right \rangle_\mc{B}^\text{asy} = \svbsf \frac{\GB \left( Br_b + Br_{\bb} \right)}{\GB + \left \langle \Gamma_\text{ion} \right \rangle}  \, . \label{eq:BSFasy}
\end{align}
Finally, bound state mediated $bb \leftrightarrow \bb \bb$ and $bb/\bb \bb \leftrightarrow VV$ scatterings constitute additional washout modes \footnote{We note that bound states can also induce a washout term
$\
   \epsilon^2 \frac{Y_\mc{B}^\text{eq}}{s}
    \GB  \frac{\GB \left( Br_b + Br_{\bb} \right)^2}{ \GB + \left \langle \Gamma_\text{ion} \right \rangle}. $
However, since it is suppressed by two powers of $\epsilon$, we can safely neglect it.}. 
In Fig.~\ref{fig:Scat_comp1}, we show the results by numerically integrating the Boltzmann equations in Eq.~\eqref{eq:BoltzmannDeltaScat}. 
We have chosen the benchmark values $\epsilon=0.1$, $m_\phi=10^4 \, \mathrm{GeV}$, $\lambda_1 = 0.1$, vary $\alpha \in \left \lbrace 0.01, 0.2 \right \rbrace$ and $z_\text{w} = \frac{m_\phi}{m_b} \ln \left[ \frac{9.5 \cdot 10^{13} \lambda_2^2}{m_b [\mathrm{GeV}]} \right] \in \left \lbrace 5,250 \right \rbrace$\footnote{
For illustrative purposes we choose the parameterization $\log_{10} \lambda_2 = 6.4 \cdot 10^{-5} z_\text{w}^{-4.19}$ and vary $\frac{m_b}{m_\phi}$ correspondingly.}
The latter estimates the time at which $\lambda_2$ mediated washouts become inefficient.
%%%%%%%%%%%%%%%%%%%%%%%%%%
%%%%%%%%%%%%%%%%%%%%%%%%%%
Contrary to the decay scenario, the ratio $m_b / m_\phi$ plays an important role, as it determines whether washouts can be efficient at times long after the freeze out of $\phi$-annihilations at $z_\text{ann}$. 
If $m_b \sim m_\phi$, the washouts typically freeze out at the time scale $z_\text{w} \lesssim z_\text{ann}$, defining a \emph{weak-washout} scenario.
Conversely, if $m_b \ll m_\phi$, washouts can still be efficient at later times, $z_\text{w} \gg z_\text{ann}$, which we call a \emph{strong-washout} scenario.
The presence of long-range interactions alters the asymmetry generation in distinct ways in these two scenarios as we discuss in the following.

\textbf{Strong washout scenario:}
The asymmetry in this scenario can be estimated as
\begin{align}
Y_{\Delta b} \sim \epsilon \eta \left[ Y_\phi \left( z_\text{w} \right) - Y_\phi \left( \infty \right) \right] \, , \\
\text{with} \quad \eta = \frac{\left \langle \sigma v \right \rangle^\text{tot}_{\scriptscriptstyle \phi \phi \rightarrow bb} + \left \langle \sigma v \right \rangle_\mc{B}^\text{asy} }{\left \langle \sigma v \right \rangle^\text{tot}_{\scriptscriptstyle \phi \phi \rightarrow bb} +  \left \langle \sigma v \right \rangle_{\scriptscriptstyle \phi \phi \rightarrow VV}^\text{eff}} \, .
\end{align}
Long-range interactions influence both the number of $\phi$ annihilating for $z>z_\text{w}$ and the efficiency $\eta$ at which these annihilations produce an asymmetry. 
The Sommerfeld effect tends to increase the asymmetry if the enhancement of the BNV channel $\phi \phi \rightarrow b b,\bb\bb$ is stronger than for the $B$-conserving annihilation $\phi \phi \rightarrow VV$, explaining the increase by up to $\mc{O} (1)$ factors in the asymmetry for relatively small $\alpha$ \footnote{This is the case for our scenario since we choose $b$ to be a $SU(N)$ singlet and thus $\phi \phi \rightarrow b b$ annihilations, in contrast to gauge annihilations, are maximally enhanced.}.
However, for large $\alpha$, BSF becomes relevant and generally decreases the asymmetry mainly due to a more efficient depletion of $\phi$ for $z<z_\text{w}$.
Then BSF outweighs the Sommerfeld effect and can lead to a net reduction of the asymmetry by up to $\sim 70 \%$.
The only instance in which bound states increase the asymmetry occurs if washouts freeze out after the asymmetry generation via bound states (cf. Eq.~\eqref{eq:BSFasy}) but before the depletion of $\phi$ into gauge bosons via bound states becomes efficient. 
This hierarchy can arise if the $B$-violating $\phi$ annihilations contain a larger singlet component of the $\phi$ scattering state than the $B$-conserving (gauge) annihilations. 
In this case, the asymmetry can be increased by up to $500 \%$.
Still, we found that this enhancement is typically not sufficient to overcome the underproduction of the BAU in the strong washout scenario.
\textbf{Weak washout scenario}: For $z_\text{w} \lesssim z_\text{ann}$, most of the asymmetry is created directly before the freeze-out of perturbative $\phi$ annihilations. 
While the long-range effects also keep the $\phi$ closer to equilibrium, their main effect is to enhance the efficiency of asymmetry generation.
This leads to a net increase of the asymmetry of up to $\sim \mc{O} (100 \%)$, which is dominated by the Sommerfeld effect, as BSF is subleading for $z \lesssim z_\text{ann}$.
For even smaller $z_\text{w} \lesssim \mc{O} \left( 1 \right)$, inverse $\phi \phi \rightarrow bb$ annihilations constitute the dominant washout contribution instead of $\lambda_2$-mediated washouts.
The inverse $\phi$ annihilations are also Sommerfeld enhanced and in addition to bound state mediated washout modes (that are resonantly enhanced for $z \lesssim \mathcal{O} \left( \frac{m_\phi}{m_\phi -m_b} \right)$) can result in the reduction of asymmetry by $\mc{O} \left( 10 \% \right)$.
Finally, the bound state and $\lambda_2$ mediated washouts can destructively interfere, which can lead to an increase of the asymmetry by a factor of a few. 
However, these cancellations only occur for specific parameter configurations, which is discussed in the appendix, and are not relevant if $\lambda_1$ is chosen sufficiently small, such as in our example.

\textbf{Discussion---}  
The presence of bound states can influence baryogenesis in three ways:
\begin{itemize}
    \item The formation of bound states that subsequently decay into gauge bosons leads to a more efficient depletion of the asymmetry generating particles and a closer to equilibrium evolution.
    \item If the bound states have $CP$- and $(B-L)$-violating decay modes, bound state formation with a subsequent $\Delta (B-L)$-decay can contribute to the creation of the asymmetry.
    \item The existence of $(B-L)$-violating decay modes leads to bound state mediated washout modes that contribute to the existing washout processes or destructively interferes with them.
\end{itemize}
We have discussed the example of heavy decaying or annihilating states in the adjoint representation of a non-abelian gauge group $SU(N_D)$. Similar results are expected to hold for other representations~\footnote{See the supplementary material for more discussion on this.}.
 
In \emph{decay dominated baryogenesis}, bound state effects can reduce the generated asymmetry and shift the viable mass range for the decaying particle by over an order of magnitude, with larger shifts for stronger gauge interactions. This situation resembles the effect of BSF for DM produced via the Super-WIMP mechanism. Two UV model examples where these results directly apply are type-II~\cite{Ma:1998dx,Lazarides:1998iq,Hambye:2000ui,Hambye:2003ka,Hambye:2005tk, GonzalezFelipe:2013jkc, AristizabalSierra:2014nzr} and type-III~\cite{Hambye:2003rt,Fischler:2008xm,AristizabalSierra:2010mv,Hambye:2012fh} seesaw models. 
Those models generate asymmetries from the decays of $SU(2)_L$ triplets and are subject to corrections due to bound states. 
If the triplets decay early  ($x_\text{dec} \lesssim 10^3$), then the relative small $SU(2)_L$ gauge coupling ($\alpha \sim 0.03$) implies a reduction of the asymmetry by $\lesssim \mc{O} (10\,\%)$. 
However, if the triplets decay relatively late ($x_\text{dec} \gtrsim 10^3$), BSF can increase the required mass of the parent particle for successful leptogenesis by a $\mc{O} \left( 1 \right)$ factor. 
For models featuring strongly interacting decaying particles ($\alpha \gtrsim 0.1$), e.g. charged under $SU(3)_C$, the required mass of the decaying particle for successful baryogensis can increase by an order of magnitude for late-decay scenarios~\cite{Babu:2012iv,Dhuria:2015xua,Kane:2019nes,Allahverdi:2017edd,Calibbi:2017rab,Gu:2017cgp,Grojean:2018fus,Fridell:2021gag}.

For \emph{scattering dominated baryogenesis}, we found that long-range interactions significantly affect the generated asymmetry if washout processes are efficient until after the freeze out of asymmetry-generating annihilations (strong washout scenario). 
In this instance, they severely alter the predicted asymmetry and increase (decrease) it by a $\mc{O} \left( 1 \right)$ factor for gauge couplings $\alpha \sim 0.01$ ($\alpha \sim 0.1$), as $(B-L)$-violating bound state decays can constitute the dominant asymmetry production channel and $(B-L)$-conserving bound state decays can efficiently deplete the particles sourcing the asymmetry. 
However, such scenarios typically feature an underproduction of the observed baryon asymmetry. 
If washout processes become inefficient before the time of $\phi$ freeze out (weak washout scenario), the long-range interactions can efficiently increase the efficiency of the asymmetry production or washouts. 
For WIMPy baryogenesis models~\cite{Cui:2011ab}, this implies that in the strong washout scenario sizeable BSF effects can occur. However, the BAU is typically underproduced due to long-lasting washouts or too large DM masses are required, overclosing the Universe. 
In the weak washout scenario instead, bound states can introduce efficient bound state mediated washout contributions that can change the asymmetry by $ \mc{O} \left( 1\%-100 \% \right)$.

In conclusion, we demonstrated the impact of bound state effects on baryogenesis mechanisms and found effects ranging from an enhancement of the predicted BAU by a factor of a few to a reduction by over an order of magnitude depending on the scenario.

\begin{acknowledgments}
\textbf{Acknowledgments---} The authors acknowledge support from the Emmy Noether grant "Baryogenesis, Dark Matter and Neutrinos: Comprehensive analyses and accurate methods in particle cosmology" (HA 8555/1-1, Project No.~400234416) funded by the Deutsche Forschungsgemeinschaft (DFG, German Research Foundation).
Moreover, M.~B.\ and J.~H.\ acknowledge support by the Cluster of Excellence “Precision Physics, Fundamental Interactions, and
Structure of Matter” (PRISMA$^+$ EXC 2118/1) funded by the Deutsche Forschungsgemeinschaft (DFG, German Research
Foundation) within
the German Excellence Strategy (Project No. 390831469). K.~F.\ acknowledges support from the Japan Society for the Promotion of Science (JSPS)
Grant-in-Aid for Scientific Research B (No. 21H01086 and 23K20847). C.~H.\ is supported by the Generalitat Valenciana under Plan Gen-T via CDEIGENT grant No. CIDEIG/2022/16.  C.~H.\ also acknowledges partial support from the Spanish grants PID2020- 113775GBI00 (AEI/10.13039/501100011033), and Prometeo
CIPROM/2021/054 (Generalitat Valenciana).
\end{acknowledgments}

\begin{center}
   \textbf{Supplementary material} 
\end{center}
\appendix

\section{\label{sec:AppA}Bound States}
In this appendix, we briefly review how to calculate the BSF cross section and furthermore illustrate how we model washout processes mediated bound states. 
%%%%%%%%%%%%%%%%%%%%%%%%%%%%%%%%%%%%%%%%%%%%%
\subsection{Bound State Formation and Decay}
Our calculation of the BSF cross section and decay rate follows Refs.~\cite{Petraki:2015hla,Harz:2018csl}. 
In the following, we summarize the results for a bound state formed from two particles in the adjoint representation of $SU(N_D)$.
Such a system can accommodate bound states transforming as singlets as well as adjoint representations of $SU(N_D)$.
In this article, we only consider singlet bound states in the ground state.
The bound state is described by the potential
\begin{align}
    V_B \left( r \right) = - k_B \frac{\alpha}{r}  \, ,
\end{align}
where $k_B=N_D$ for a singlet bound state. 
We have assumed massless gauge bosons, and where $\alpha$ is the fine structure constant of the $SU(N_D)$. 
Since the formation of a singlet bound state, $XX \rightarrow \mc{B} V$, proceeds via the emission of a gauge boson $V$, which is in the adjoint representation, the initial state $XX$ has to also be in the adjoint representation. 
The scattering state is then described by the potential
\begin{align}
    V_S \left( r \right) = -k_S \frac{\alpha}{r} \, ,
\end{align}
where $k_S = \frac{N_D}{2}$ for a scattering state in the adjoint representation.
The BSF cross section can then be calculated by evaluating an overlap integral of the scattering state and bound state wave function
\begin{align}
    \sigma_\mc{B}^{\mathbf{adj} \rightarrow \mathbf{sing}} v = \frac{2^9 \pi}{3} \frac{\alpha^\text{BSF} \alpha_g^B}{m_X^2} \frac{ N_D^3 + 4 N_D}{8\left( N_D^2 + 1 \right)^2} S_\mc{B} \left( \frac{\alpha_g^S}{v}, \frac{\alpha_g^B}{v} \right)  \, . \label{eq:BSFCS_APP}
\end{align}
The superscripts $B$ and $\text{BSF}$ for $\alpha$ indicate at which energy scale the interaction strength has to be evaluated, where $\alpha_g^{S/B} = k_{S/B} \alpha$, see discussion around Eq.~\eqref{eq:BSF_CS}. 
For simplicity we do not consider the running of the gauge coupling allowing for BSF and refer the reader to \cite{Harz:2018csl} for details about the precise treatment of this issue and the estimated size of the induced corrections. 
The factor 
\begin{equation}
\begin{aligned}
    S_\text{BSF} \left( \zeta_S, \zeta_B \right) &= \frac{2 \pi \zeta_s}{1 - \exp(- 2 \pi \zeta_S)} (1 + \zeta_S^2) \\
    &\quad\times \frac{\zeta_B^4 \exp \left( - 4 \zeta_S \arccot \left( \zeta_B \right) \right)}{\left( 1+\zeta_B^2 \right)^3} \, \label{eq:SommerfeldFactor}
\end{aligned}
\end{equation}
describes the Sommerfeld factor, where $\zeta_{S/B} = k_{S/B} \, \alpha/v$. 
The thermally averaged BSF cross section $\svbsf$ is evaluated in the non-relativistic limit, and the bound state ionization (or dissociation) rate $\Gamma_\text{ion}$ can be related to the thermally averaged BSF cross section using the Milne relation and non-relativistic approximations for the equlibriums yields involved, such that
\begin{align}
   \left \langle \Gamma_\text{ion} \right \rangle = \left \langle \sigma_\mc{B} v \right \rangle s \frac{\left( Y_X^\text{eq} \right)^2}{Y_\mc{B}^\text{eq}} \, .
\end{align}
The bound state decay rate can be approximately related to the $s$-wave contribution of the corresponding annihilation cross section of the bound states constituents into $\mc{F}$ final states,
\begin{align}
    \Gamma_{\mc{B} \rightarrow \mc{F}} = |\psi \left( 0 \right)|^2 \left( \sigma \left( XX \rightarrow \mc{F} \right)^\text{s-wave}_\text{sing.} v \right) \, , \label{eq:BDecayRate}
\end{align}
where $\psi (0) = {m_X^3 k_B^3 \alpha^3}/{8 \pi}$ is the ground state wave function at the origin. 
This implies that if the constituent annihilations violate baryon- or lepton number, the bound state decays will also violate baryon- or lepton number. 
%%%%%%%%%%%%%%%%%%%%%%%%%%%%%%%%%%%%
\subsection{Bound State Formation vs.~Perturbative Annihilation}
The existence of bound states can only influence the generation of a baryon asymmetry if the bound states are formed efficiently. 
For baryogenesis via decays, they impact the final asymmetry by more efficiently reducing the number density of particles that source the asymmetry. 
This happens if BSF sizeably alters the effective annihilation cross section (see Eq.~\eqref{eq:effective_annihilation}) at the relevant time scale. 

\begin{figure}
    \centering
    \includegraphics[width=0.98\columnwidth]{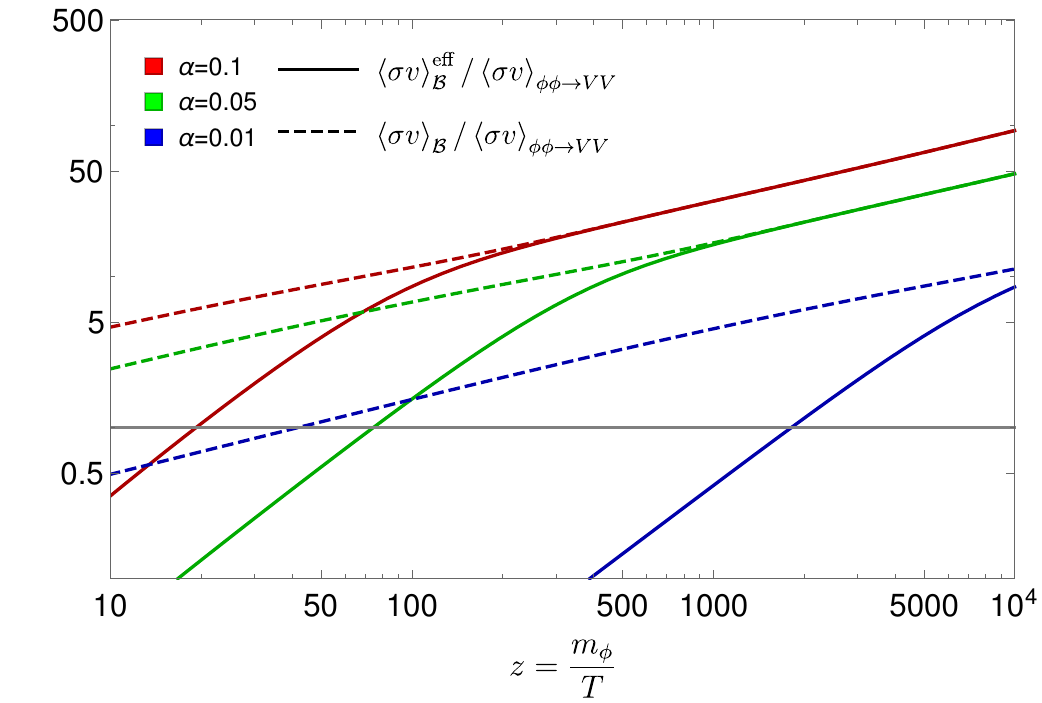}
    \caption{The time evolution of the ratio of the effective annihilation cross section via BSF and subsequent bound state decay $\left \langle \sigma v \right  \rangle_\mc{B}^\text{eff}$ and the perturbative annihilation cross section $\left\langle \sigma v \right\rangle_{\phi \phi \rightarrow VV}$ is shown in solid lines, while the ratio of the BSF cross section $\svbsf$ and the perturbative annihilation cross section $\left\langle \sigma v \right\rangle_{\phi \phi \rightarrow VV}$ is depicted in dashed lines. The color code indicates the size of the gauge coupling considered and the solid gray line indicates where two compared quantities are equal.}
    \label{fig:BSFvsPert}
\end{figure}

In Fig.~\ref{fig:BSFvsPert} we illustrate the ratio of the effective annihilation cross section induced by BSF and subsequent decay $\left \langle \sigma v \right \rangle_\mc{B}^\text{eff}$ (see Eq.~\eqref{eq:effective_annihilation}) and the perturbative annihilation cross section $\left \langle \sigma v \right \rangle_{\scriptscriptstyle XX \rightarrow VV}$. 
%%%%%%%%%%%%%%%%%%%%%%%%%%%%%%%%%%%%%%%%%%
We see that annihiliation via BSF dominates over the perturbative annihilations at late times. 
When BSF dominates depends on the size of the gauge coupling, since bound states are not efficiently ionized anymore as soon as the temperature drops below the binding energy $T \lesssim E_\mc{B} \sim m_X \alpha^2$. 
As a rule of thumb, we find that BSF is dominant for $5 x \alpha^2  \gtrsim 1$. 
This estimate follows from the requirement $T \lesssim E_B$ and the numerical prefactor reflects our numerical findings.

\begin{figure}
    \centering
    \includegraphics[width=0.98\columnwidth]{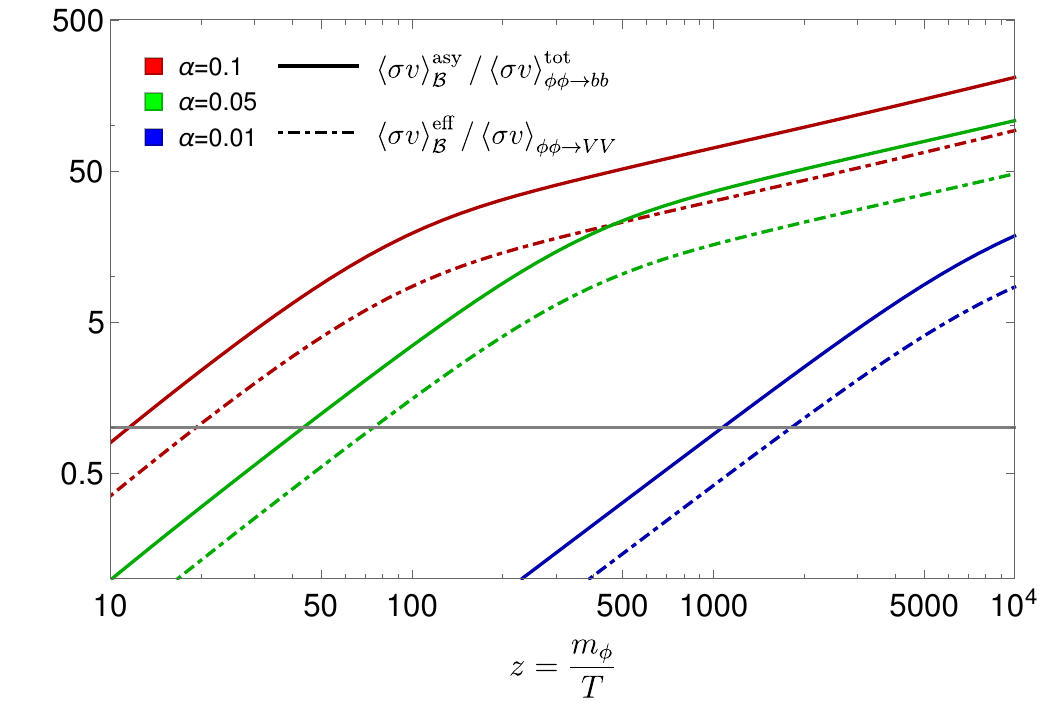}
    \caption{The time evolution of the ratio between the bound state induced asymmetry generating annihilation cross section $\left \langle \sigma v \right \rangle_\mc{B}^\text{asy}$ (annihilation cross section $\left \langle  \sigma v \right\rangle_\mc{B}^\text{eff}$) and the corresponding perturbative cross section $\left \langle  \sigma v \right\rangle_{\phi \phi \rightarrow bb}$ ($\left\langle \sigma v \right\rangle_{\phi \phi \rightarrow VV}$) for various gauge couplings $\alpha$ are shown in solid (dot-dashed) lines. The color indicates the size of the gauge coupling $\alpha$, the solid gray line indicates where the compared cross sections are equal and we have set $\lambda_1=0.1$. }
    \label{fig:BSFvsAnnAsy}
\end{figure}

For {\it scattering dominated baryogenesis}, BSF with a subsequent baryon number violating decay can generate a baryon asymmetry. 
In Fig.~\ref{fig:BSFvsAnnAsy}, we compare this asymmetry generating effective cross section $\left \langle \sigma v \right\rangle_\mc{B}^\text{asy}$ [see Eq.~\eqref{eq:BSFasy}] to the perturbative asymmetry generating rate $\left \langle \sigma v \right \rangle^\text{tot}_{\phi \phi \rightarrow bb}$. 
We as well show the effective bound state induced annihilation cross section $\left \langle \sigma v \right \rangle_\mc{B}^\text{eff}$ compared to its perturbative counterpart $\left \langle \sigma v \right \rangle_{\scriptscriptstyle XX \rightarrow VV}$.
The behavior is similar to the effective dark sector annihilation cross section discussed before. 
The earlier rise of the asymmetry generating bound state induced contribution originates from the larger singlet component of the asymmetry generating $\phi$ annihilations, which is chosen to be maximal, compared to the singlet component of the $\phi$ gauge annihilations.
%%%%%%%%%%%%%%%%%%%%%%%%%%%%%%%%%%%%%%%%%%%%%%%%%

At this point we would like to comment on the implications of choosing representations of the bound state constituents different from the adjoint. 
A different representation induces a different group factor in the BSF cross section [the $N_D$-dependent factor in Eq.~\eqref{eq:BSFCS_APP}], but we expect the most significant change to come from the Sommerfeld factor [see Eq.~\eqref{eq:SommerfeldFactor}]. 
For two particles in the adjoint, the potential of the scattering state in the BSF process is attractive ($k_S>0$), leading to an increase of the thermally averaged cross section proportional to $\sqrt{z}$, as can be seen in Figs.~\ref{fig:BSFvsPert} and \ref{fig:BSFvsAnnAsy}. 
However, BSF can also proceed from repulsive scattering states ($k_S<0$), as for example for BSF from particles in the (anti-)fundamental representation. 
In this scenario, BSF is exponentially suppressed at large $z$. 
For instance, for a particle-anti-particle bound state where the particle transforms under the fundamental representation of $SU(3)_C$, this suppression would start to be relevant for $z \gtrsim 10^3$, and BSF would be inefficient~\cite{Harz:2018csl}.
In decay dominated baryogenesis, we would expect similar effects from BSF for all scenarios where the asymmetry is generated at $z \gtrsim 10^3$, for $X$ being in the fundamental representation.
This is different from the scenario discussed in the main text, since for $X$ in the adjoint representation, the BSF cross section is unsuppressed even for $z \gtrsim 10^3$.  

Similarly, the consideration of excited bound states will also alter the BSF cross section. 
Excited bound states have a smaller binding energy and start to become relevant at later times. 
We expect them to be of particular relevance if the dynamics generating or depleting the asymmetry take place at very late times where they could significantly alter the results, increasing the impact of bound states. 
This has been demonstrated in the context of SuperWIMP dark matter production~\cite{Binder:2023ckj}.

%%%%%%%%%%%%%%%%%%%%%%
\begin{figure}
    \centering
    \includegraphics[width=\columnwidth]{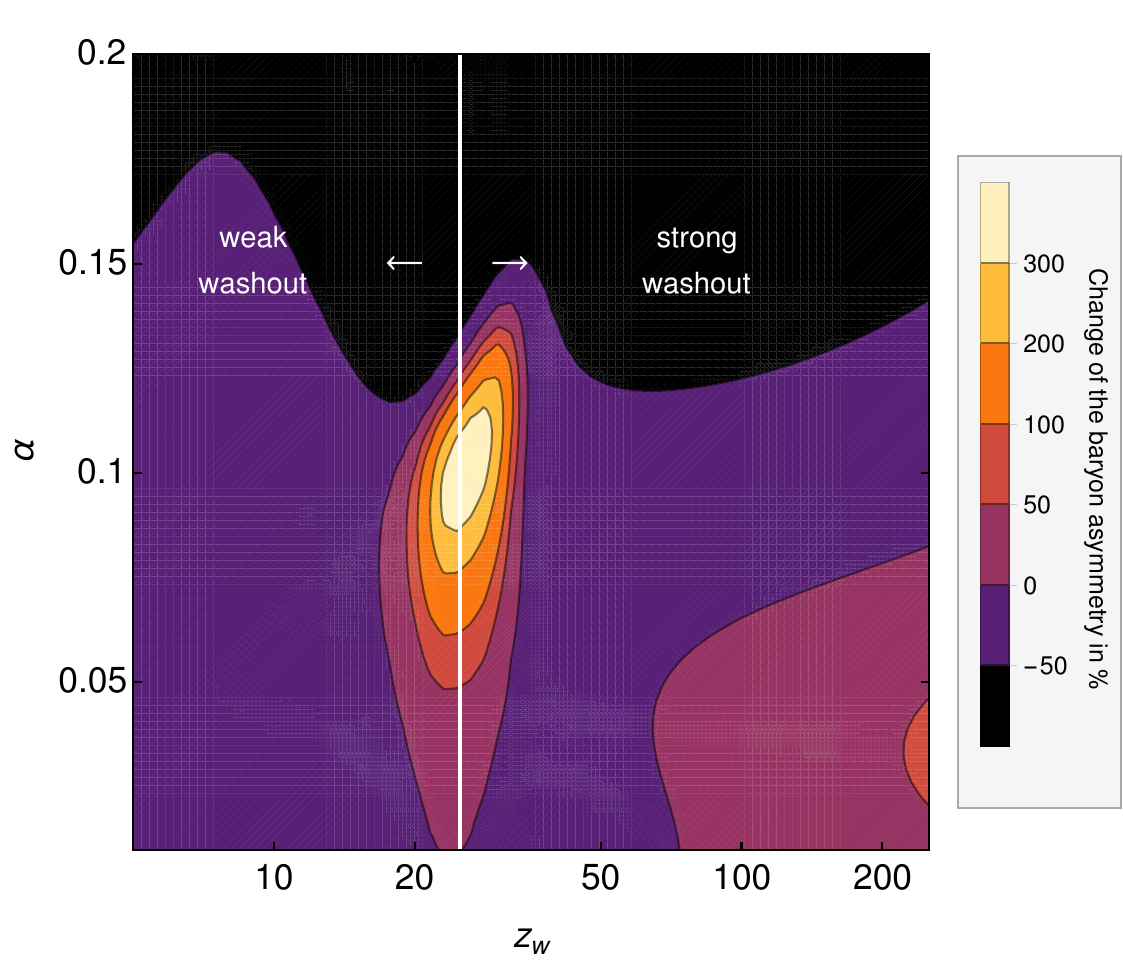}
    \caption{{\bf Scattering dominated baryogenesis scenario.} The change of the generated asymmetry in $\%$ in the $z_\text{w}$-$\alpha$ plane for $m_\phi=10^4 \, \mathrm{GeV}$, $\lambda_1=3$. The white line indicates the transition from the weak to the strong washout regime.}
    \label{fig:Scat_comp2}
\end{figure}
%%%%%%%%%%%%%%%%%%%%%%%%%%

%%%%%%%%%%%%%%%%%%%%%%%%%%%%%%%%%%%
\subsection{Bound State Mediated Washout}
In the scattering dominated baryogenesis scenario discussed in the main text, bound states $\mathcal{B}$ can decay into final states carrying non-zero $(B-L)$, specifically $\mathcal{B} \rightarrow bb$ or $\mathcal{B} \rightarrow \bb \bb$.
In analogy to decay dominated baryogenesis, baryon-number violating scatterings $bb \leftrightarrow \bb \bb$ mediated by bound states have to be considered in addition to $bb \leftrightarrow \bb \bb$ scatterings mediated by $\lambda_2$ [see Eqs.~\eqref{eq:l2scattering1} and \eqref{eq:l2scattering2}].   
We calculate the cross section for this process in the following way:
We deduce a coupling constant $\lambda_{\mc{B} bb}$ of an interaction vertex $\mc{L} \supset \lambda_{\mc{B} bb} \mc{B} bb$ from the decay rate $\Gamma \left( \mathcal{B} \rightarrow bb \right)$ given by Eq.~\eqref{eq:BDecayRate}, and find
\begin{align}
    \lambda_{\mc{B} bb} = \sqrt{\frac{k_B^3 \alpha^3}{2^5 \pi}}\lambda_1 m_\mc{B} \sqrt{1+\epsilon} \, . \label{eq:lambdaboundstate}
\end{align}
The expression for $\lambda_{\mc{B} \bb \bb}$ is the same but exchanging $\epsilon \rightarrow - \epsilon$. \\
We calculate the cross section of $bb \leftrightarrow \bb \bb$ considering both the contact interaction vertex and the bound state mediated interaction resulting from the interaction Lagrangian $\mc{L} \supset \lambda_2 bb \bb \bb + \lambda_{\mc{B} bb} \mc{B} bb + \lambda_{\mc{B} \bb \bb} \mc{B} \bb \bb$, leading to
\begin{equation}
\begin{aligned}
    \sigma \left( bb \leftrightarrow \bb \bb \right) =& \frac{1}{16 \pi s}\times\\
    &\!\!\!\!\!\!\!\!\!\!\!\!\!\!\!\!\!\left[ \lambda_2^2 + \frac{2\lambda_2 \lambda_{\mc{B} bb} \lambda_{\mc{B} \bb \bb} \left( s - m_\mc{B}^2 \right) + \lambda_{\mc{B} b b}^2 \lambda_{\mc{B} \bb \bb}^2 }{\left( s - m_\mc{B}^2 \right)^2 + \Gamma_\mc{B}^2 m_\mc{B}^2} \right]  \, , \label{eq:BmedWash}
\end{aligned}
\end{equation}
where we have neglected contributions of $\mc{O} (\epsilon^2)$.
The cross section for $VV \leftrightarrow bb/\bb \bb$, mediated by a bound state in the s-channel, can be obtained in an analogous way from the operator $\lambda_{\mc{B} V V} \mc{B} G_a^{\mu \nu} G^a_{\mu \nu}$, where $G_a^{\mu \nu}$ is the field strength tensor of the $SU(N_D)$. 

When $\lambda_1^2 \gg \lambda_2$, the bound state mediated washouts effectively increase the efficiency of the $bb \leftrightarrow \bb \bb$ washouts (see Eqs.~\eqref{eq:BmedWash} and \eqref{eq:lambdaboundstate}) and therefore reduce the produced asymmetry, which was discussed in the main text. 

%%%%%%%%%%%%%%%%%%%%%%%%%%
\begin{figure}[b]
    \centering
    \includegraphics[width=0.98\columnwidth]{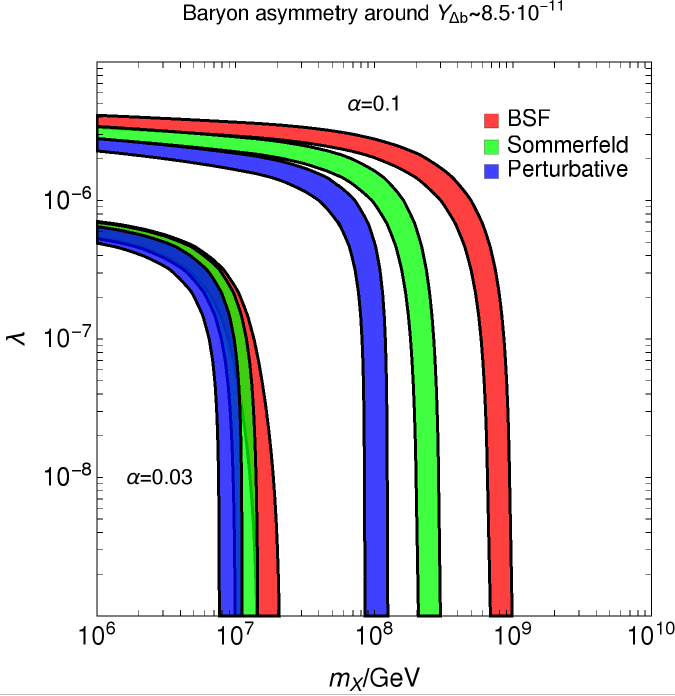}
    \caption{The region in red (green, blue) indicates where the models generates an asymmetry $Y_{\Delta b}$ in between $6.5 \leq Y_{\Delta b} \cdot 10^{-11} \leq 10.5$ including BSF (the Sommerfeld effect, perturbative annihilations only) for the decay dominated baryogenesis scenario. The three upper bands show results for a gauge coupling of $\alpha=0.1$ ($SU(3)_C$-like), while the three lower bands show results for $\alpha=0.03$ ($SU(2)_L$-like).}
    \label{fig:Decay_a01}
\end{figure}
%%%%%%%%%%%%%%%%%%%%
\begin{figure*}[t]
  \newcommand*\FigVSkip{0.5em}
  \newcommand*\FigHSkip{1.5em}
  \newsavebox\FigBox
  \begin{minipage}{\textwidth}
    \centering
  \sbox{\FigBox}{\includegraphics[width=0.4 \textwidth]{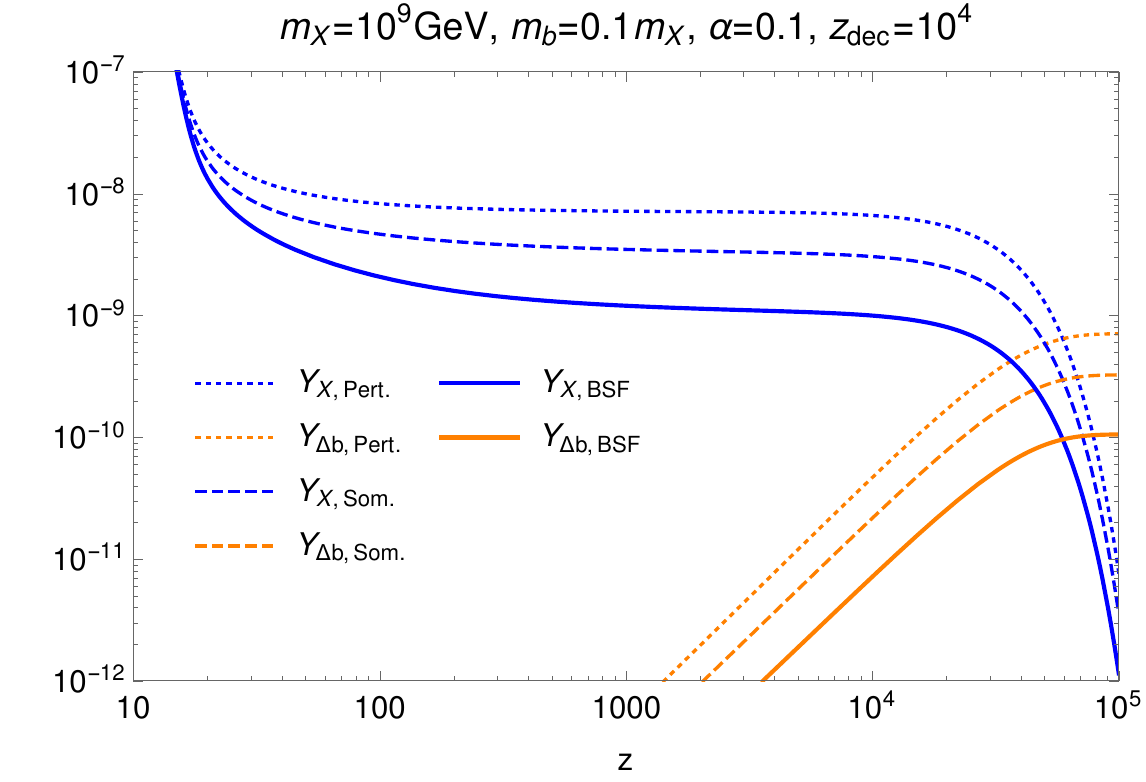}}
  \begin{minipage}{\wd\FigBox}
    \centering\usebox{\FigBox} 
   \label{fig:Decay_Nexampe_1}
  \end{minipage}\hspace*{\FigHSkip}
  % Save second image 
  \sbox{\FigBox}{\includegraphics[width=0.4 \textwidth]{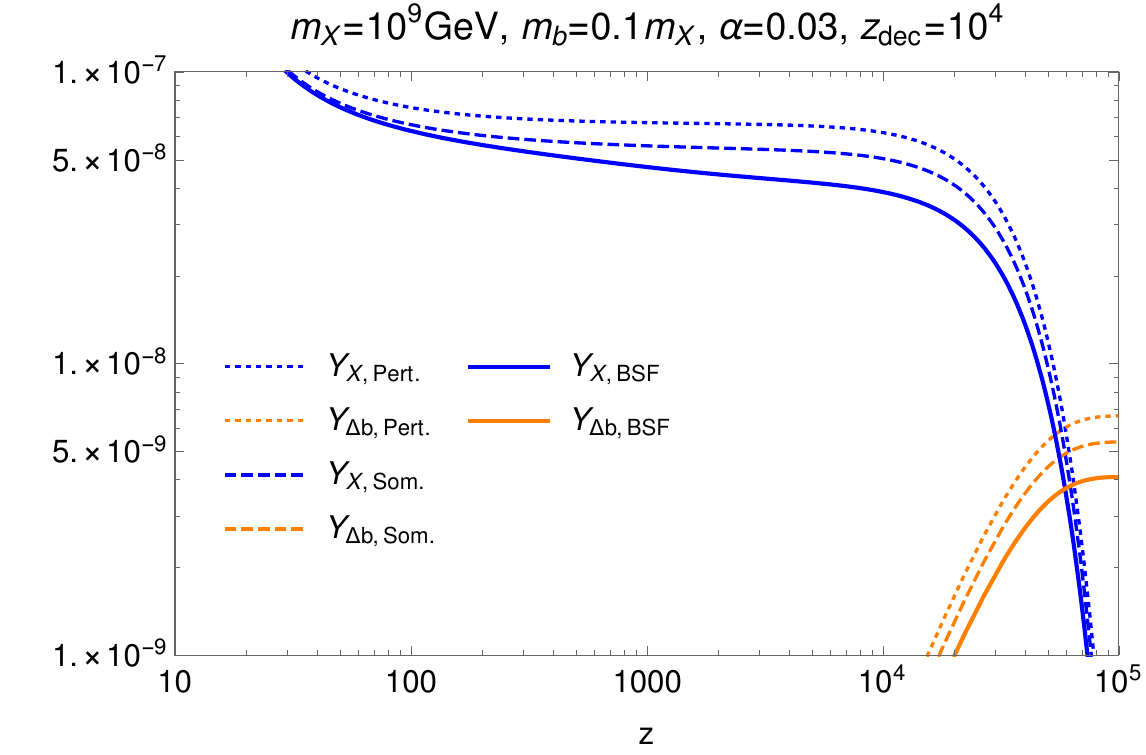}}
  \begin{minipage}{\wd\FigBox}
    \centering\usebox{\FigBox}
    \label{fig:Decay_Nexampe_2}
  \end{minipage}\\\vspace*{\FigVSkip}
   \sbox{\FigBox}{\includegraphics[width=0.4 \textwidth]{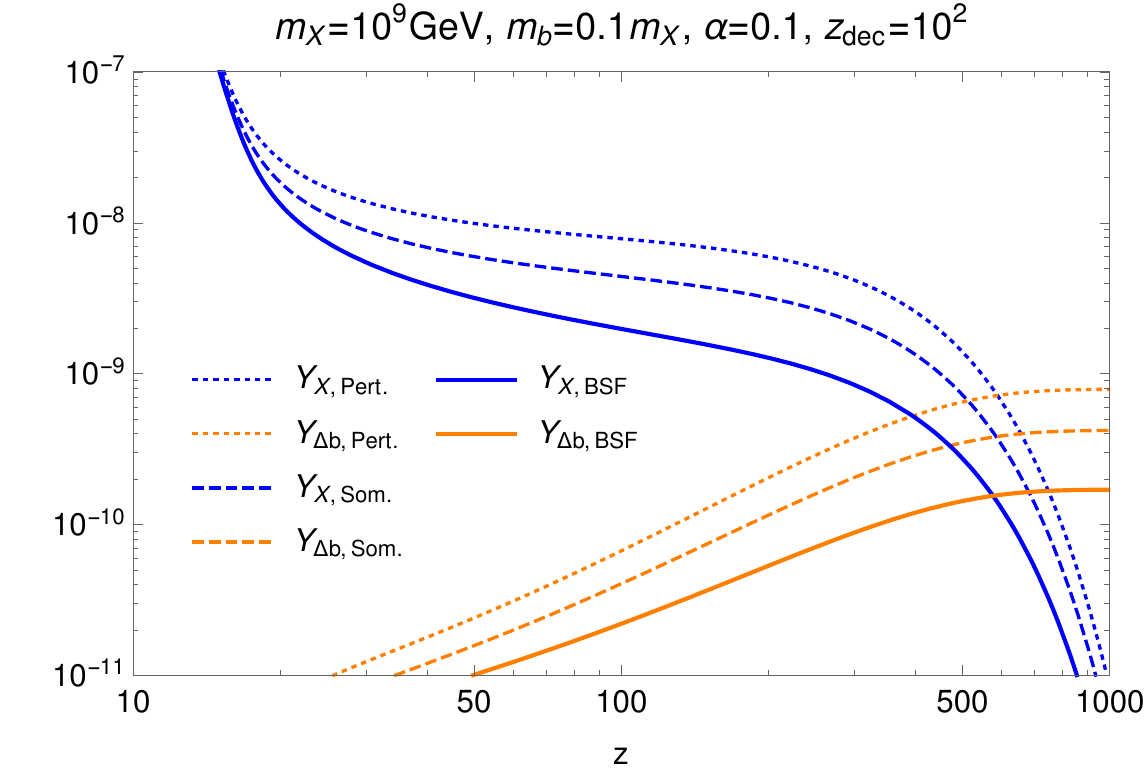}}
  \begin{minipage}{\wd\FigBox}
    \centering\usebox{\FigBox}
    \label{fig:Decay_Nexampe_3}
  \end{minipage}\hspace*{\FigHSkip}
  % Save second image 
  \sbox{\FigBox}{\includegraphics[width=0.4 \textwidth]{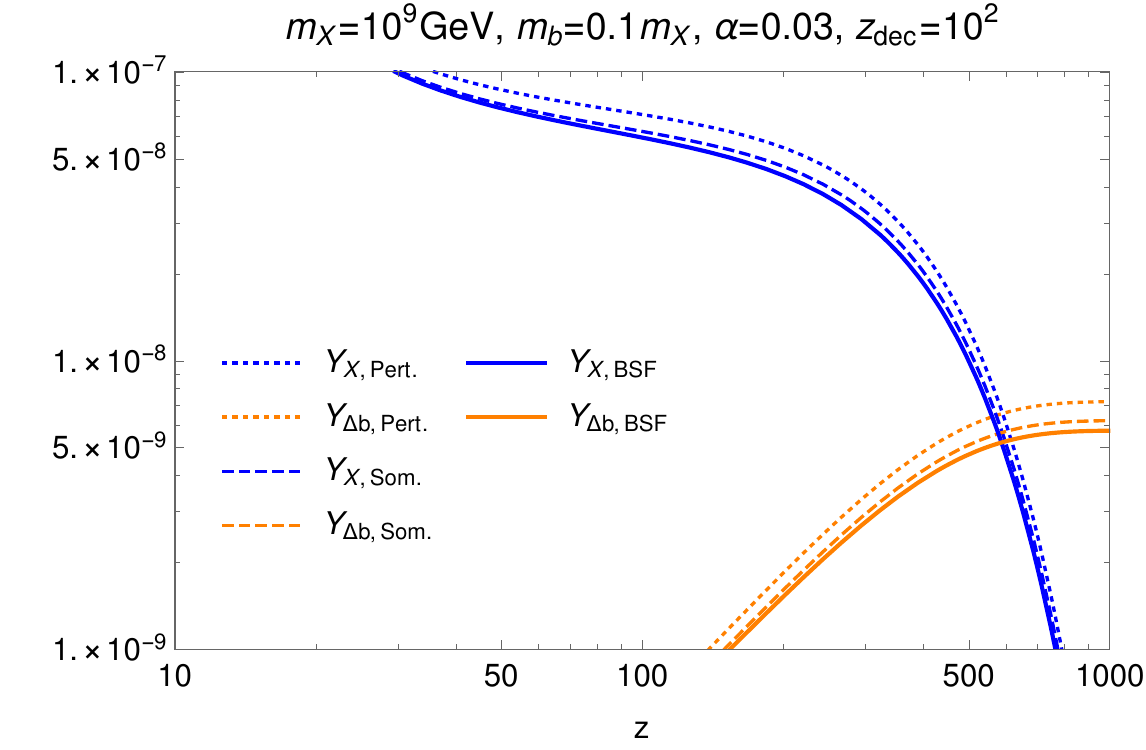}}
  \begin{minipage}{\wd\FigBox}
    \centering\usebox{\FigBox}
    \label{fig:Decay_Nexampe_4}
  \end{minipage}\\\vspace*{\FigVSkip}
    \end{minipage}
  \caption{{\bf Decay dominated baryogenesis scenario:} Density evolution with parameter choices as indicated in the figure. }
  \label{fig:DecayEvo}
\end{figure*}
\begin{figure*}[t]
  \newcommand*\FigVSkip{0.5em}
  \newcommand*\FigHSkip{1.5em}
 % \newsavebox\FigBox
  \begin{minipage}{\textwidth}
    \centering
   \sbox{\FigBox}{\includegraphics[width=0.4 \textwidth]{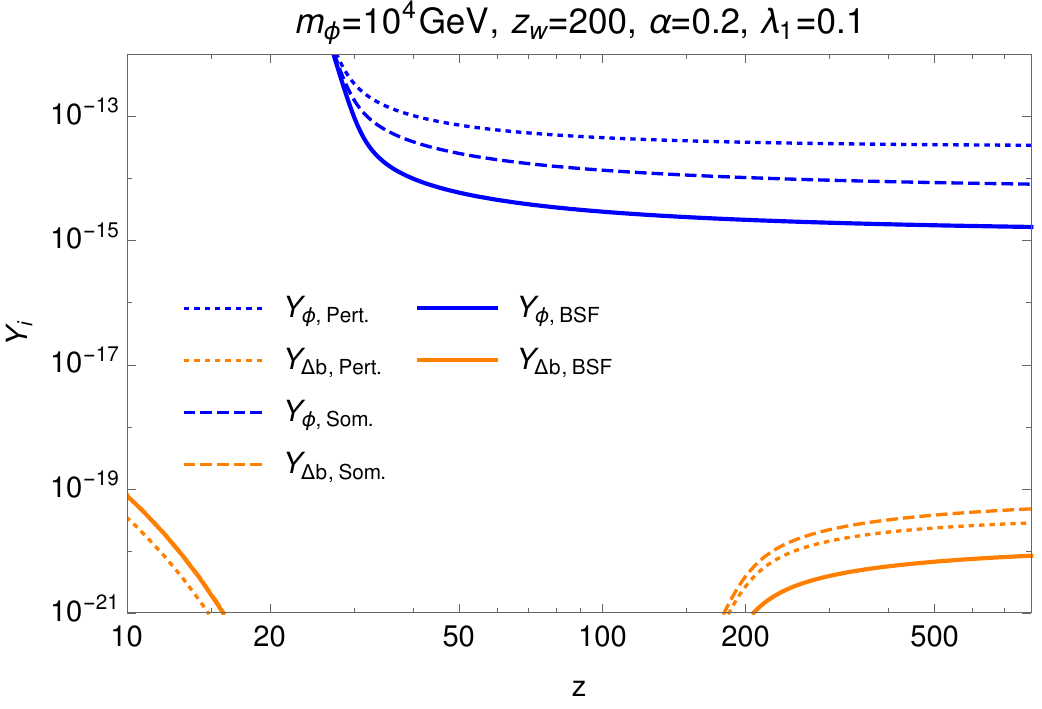}}
  \begin{minipage}{\wd\FigBox}
    \centering\usebox{\FigBox}
   \label{fig:Scattering_Nexampe_1}
  \end{minipage}\hspace*{\FigHSkip}
  % Save 3 image 
  \sbox{\FigBox}{\includegraphics[width=0.4 \textwidth]{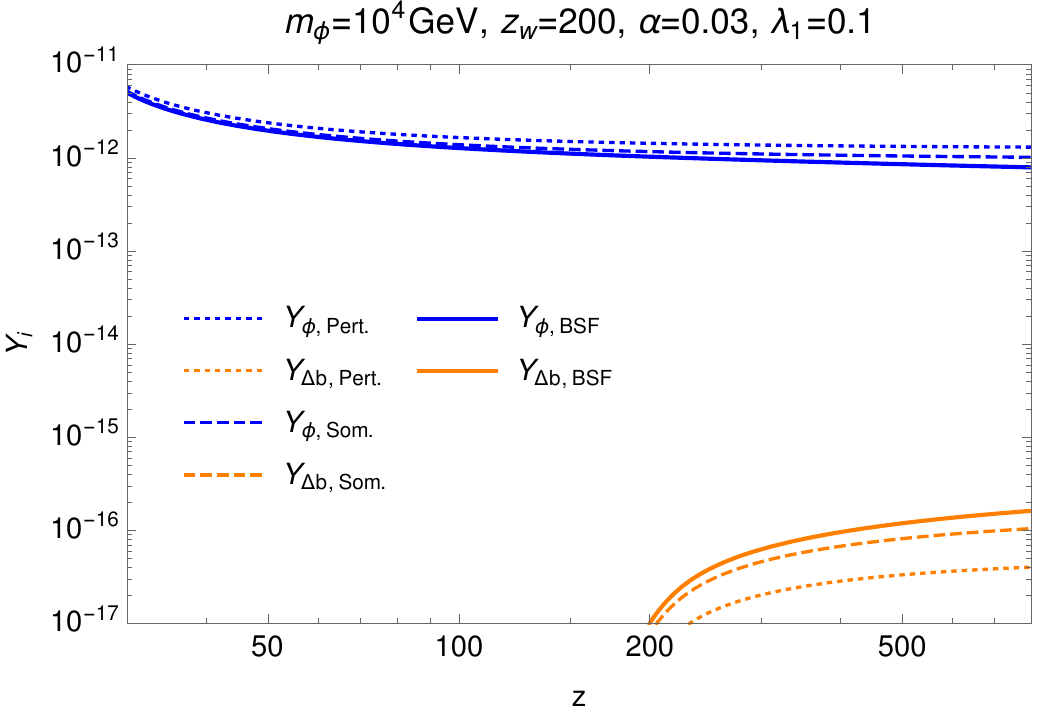}}
  \begin{minipage}{\wd\FigBox}
    \centering\usebox{\FigBox}
    \label{fig:Scattering_Nexampe_2}
  \end{minipage}\\\vspace*{\FigVSkip}
     \sbox{\FigBox}{\includegraphics[width=0.4 \textwidth]{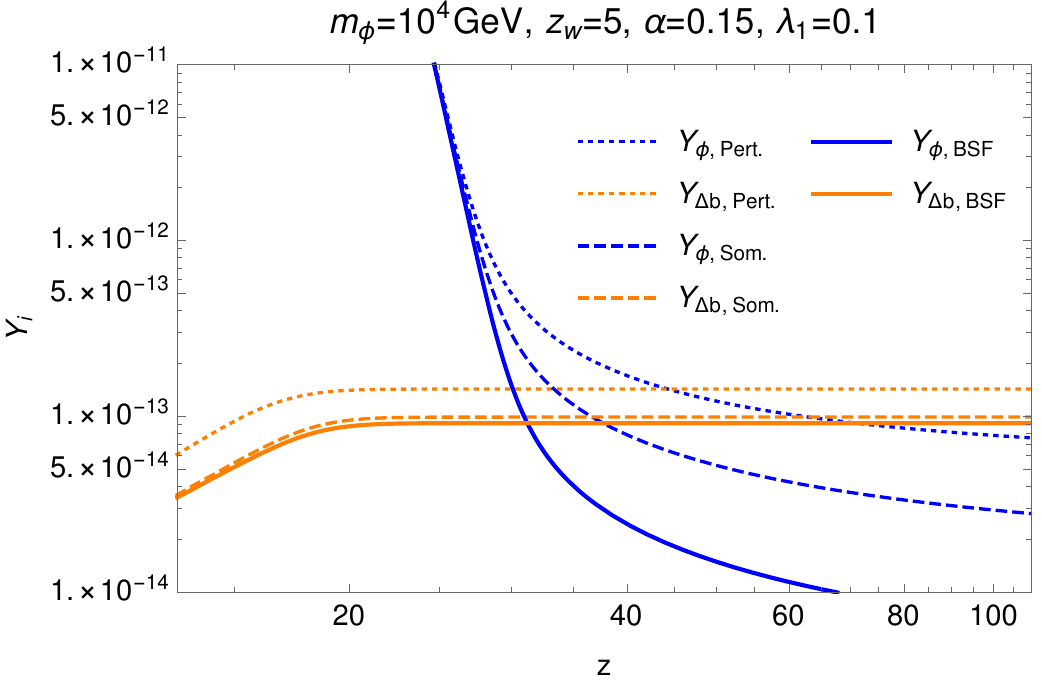}}
  \begin{minipage}{\wd\FigBox}
    \centering\usebox{\FigBox}
    \label{fig:Scattering_Nexampe_3}
  \end{minipage}\hspace*{\FigHSkip}
  % Save 4 image 
  \sbox{\FigBox}{\includegraphics[width=0.4 \textwidth]{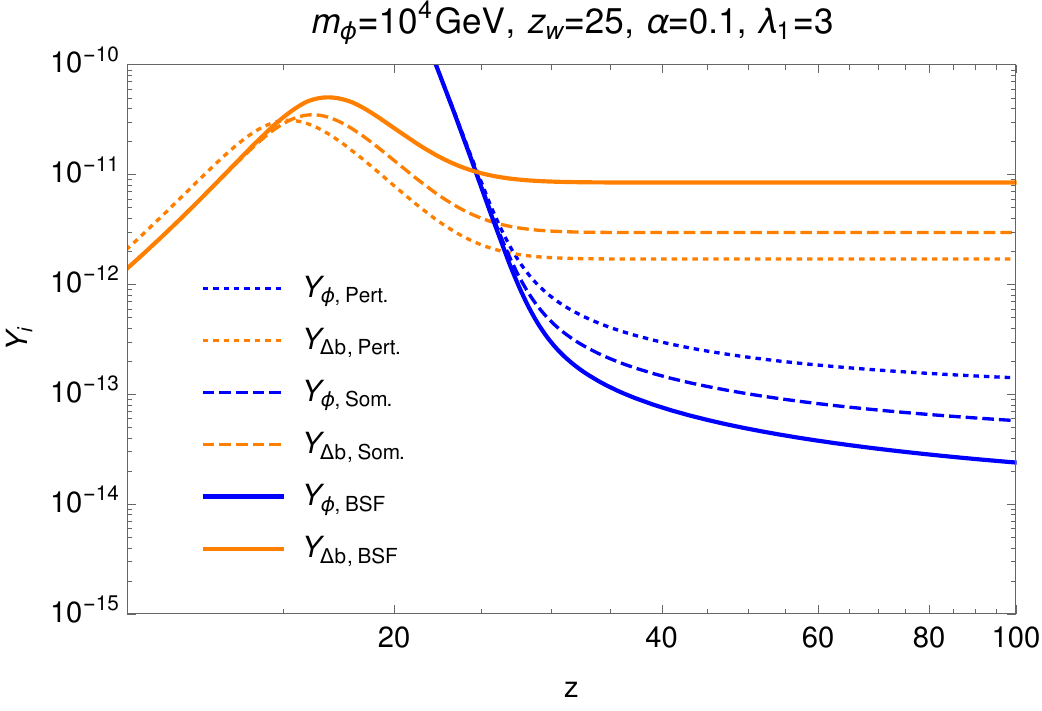}}
  \begin{minipage}{\wd\FigBox}
    \centering\usebox{\FigBox}
    \label{fig:Scattering_Nexampe_4}
  \end{minipage}
    \end{minipage}
  \caption{{\bf Scattering dominated baryogenesis scenario:} Density evolution with parameter choices as indicated in the figure. }
  \label{fig:ScattEvo}
\end{figure*}

Furthermore, the contact interaction mediated by $\lambda_2$ and the bound state mediated contribution can destructively interfere. 
Destructive interference can only occur if 
\begin{align}
    \frac{\lambda_1^2}{\lambda_2} < \frac{2^5 \pi}{k_B^3 \alpha^3} \, \oo{,}
\end{align}
and, for $m_b \ll m_\phi$, is maximal if 
\begin{align}
    \lambda_2 = \frac{3}{128 \pi} \lambda_1^2 \left( k_B \alpha \right)^3 \, .
\end{align}
Those cancellations are irrelevant for the parameter choice presented in the main text, because we choose $\lambda_1=0.1$. 
However, if we perform the same analysis with a larger coupling $\lambda_1$, for instance $\lambda_1=3$, we find an overall reduction of the asymmetry of $\mc{O} (10 \%)$ everywhere but in a region where the cancellation induced by bound state mediated washout channels reduces the efficiency of the washouts significantly.
In this region the asymmetry can be increased by $\mc{O} (100 \%)$, see Fig.~\ref{fig:Scat_comp2}.

\section{\label{sec:AppB}Impact of BSF on the Parameter Space of Successful Baryogenesis}
In the following, we illustrate the impact of bound state effects on the parameter space ($\lambda$, $m_X$) of successful baryogenesis. 
For this purpose, to mimic $SU(3)_C$ and $SU(2)_L$, we use the gauge couplings $\alpha = \left \lbrace 0.03, 0.1 \right \rbrace$, set the asymmetry parameter $\epsilon = 0.1$ and $m_b= 0.1 m_X$. We solve the Boltzmann equations corresponding to the decay-dominated scenario for various values of parent particle mass $m_X$ and asymmetry generating couplings $\lambda$.  
In Fig.~\ref{fig:Decay_a01} we show where the final baryon asymmetry is close to the experimentally observed value (assuming that the particle $b$ carries $B=1$) in the decay dominated scenario including only perturbative gauge annihilations (blue), Sommerfeld enhanced gauge annihilations (green) and BSF (red). 
We observe a shift by one order of magnitude towards larger masses $m_X$ if the asymmetry is created predominantly after the abundance of the heavy particle $X$ has been frozen out and the gauge interaction mediating BSF is relatively strong ($\alpha=0.1$).
For $\alpha=0.03$, the effect is milder and the parent particle mass corresponding to successful baryogenesis is shifted by a factor of a few. 

\section{\label{sec:AppC}Density Evolution}
In order to get an intuition of the evolution of the abundance of $X$ or $\phi$ particles and the baryon asymmetry $Y_{\Delta b}$ and the corresponding impact of the Sommerfeld effect and bound state formation, we show in the following few examples. In Fig.~\ref{fig:DecayEvo} we present the density evolution of the decay dominated scenario for $z_\text{dec} = \lbrace 10^2,10^4 \rbrace$ and $\alpha = \lbrace 0.03, 0.1 \rbrace$. 
The depletion of the heavy particle $X$ via BSF becomes efficient for $z \gtrsim 0.2 \alpha^{-2}$ (solid blue lines) and translates into a decrease of the asymmetry if $z_\text{dec}$ is sufficiently large.

In the Fig.~\ref{fig:ScattEvo}, we show exemplary density evolutions for the scattering dominated scenario with the parameters as stated in the figures. 
The upper row represents a strong-washout scenario where BSF decreases or increases the asymmetry depending on the gauge coupling. 
The lower row illustrates a weak-washout scenario, where on the left bound states are less relevant while on the right the couplings are chosen to allow for bound state caused destructive interference in the washout channels leading to an increase of the asymmetry.

\bibliographystyle{apsrev4-1}
\twocolumngrid
\bibliography{biblio} 

\end{document}